\def\useextern{}

\RequirePackage{cmap}

\documentclass[conference,final]{IEEEtran}[2015/08/26]

\usepackage{amsmath}

\usepackage[zerostyle=b,scaled=.75]{newtxtt}

\usepackage{mwe}

\usepackage{ifluatex}
\ifluatex
\else
\usepackage[T1]{fontenc}
\usepackage[utf8]{inputenc} 
\fi

\usepackage{graphicx}

\usepackage{upquote}

\usepackage{paralist}

\usepackage{fvextra}
\usepackage{csquotes}

\usepackage[final,protrusion=alltext-nott]{microtype}
\DisableLigatures{encoding = T1, family = tt*}

\usepackage{hyperref}

\usepackage{url}
\makeatletter
\g@addto@macro{\UrlBreaks}{\UrlOrds}
\makeatother

\usepackage{booktabs}

\usepackage{xcolor}

\usepackage{listings}
\usepackage{chngcntr}
\AtBeginDocument{\counterwithout{lstlisting}{section}}

\AtBeginEnvironment{listing}{\setcounter{listing}{\value{lstlisting}}}
\AtEndEnvironment{listing}{\stepcounter{lstlisting}}

\usepackage{pdfcomment}
\ifCLASSOPTIONcompsoc
\usepackage[%
square,        
comma,         
numbers,       
sort           
]{natbib}
\else
\usepackage[%
square,        
comma,         
numbers,       
sort&compress 
]{natbib}
\fi

\usepackage{etoolbox}
\makeatletter
\patchcmd{\NAT@test}{\else \NAT@nm}{\else \NAT@hyper@{\NAT@nm}}{}{}
\makeatother

\usepackage[all]{hypcap}

\usepackage[capitalise,nameinlink]{cleveref}
\crefname{lstlisting}{\lstlistingname}{\lstlistingname}
\Crefname{lstlisting}{Listing}{Listings}

\makeatletter
\@namedef{ver@lineno.sty}{9999/12/31}
\@namedef{opt@lineno.sty}{}
\makeatother

\usepackage{etoolbox}
\ifdef{\genextern}{%
\usepackage[newfloat,finalizecache]{minted}
}{%
\ifdef{\useextern}{%
\usepackage[newfloat,frozencache]{minted}
}{%
\usepackage[newfloat]{minted}
}}%

\setminted{numbersep=5pt, xleftmargin=12pt, baselinestretch=0.8}

\usepackage[labelfont=bf,font=small,skip=4pt]{caption}
\SetupFloatingEnvironment{listing}{name=Listing,within=none}
\usepackage{xspace}

\DeclareFontFamily{U}{MnSymbolC}{}
\DeclareSymbolFont{MnSyC}{U}{MnSymbolC}{m}{n}
\DeclareFontShape{U}{MnSymbolC}{m}{n}{
	<-6>    MnSymbolC5
	<6-7>   MnSymbolC6
	<7-8>   MnSymbolC7
	<8-9>   MnSymbolC8
	<9-10>  MnSymbolC9
	<10-12> MnSymbolC10
	<12->   MnSymbolC12%
}{}
\DeclareMathSymbol{\powerset}{\mathord}{MnSyC}{180}

\usepackage{flushend}

\usepackage{comment}

\usepackage{graphicx}

\usepackage{float}

\newmintinline{c}{style=bw}
\newmintinline{text}{style=bw}

\usepackage{multirow}

\usepackage{subcaption}

\crefname{sublisting}{Listing}{Listings}
\Crefname{sublisting}{Listing}{Listings}

\usepackage{minibox}

\ifdefined\useextern
\usepackage{tikzexternal}
\else
\usepackage{tikz}

\ifdefined\genextern
\usetikzlibrary{external}
\fi

\usetikzlibrary{calc}
\usetikzlibrary{decorations}
\usetikzlibrary{decorations.pathreplacing}
\usetikzlibrary{decorations.text}
\usetikzlibrary{decorations.pathmorphing}
\usetikzlibrary{decorations.markings}
\usetikzlibrary{fit}
\usetikzlibrary{arrows}    
\usetikzlibrary{shapes}
\usetikzlibrary{shadows}
\usetikzlibrary{automata}
\usetikzlibrary{positioning}
\usetikzlibrary{petri}
\usetikzlibrary{chains}
\usetikzlibrary{fadings}
\usetikzlibrary{matrix}
\usetikzlibrary{patterns}
\usetikzlibrary{mindmap}
\usetikzlibrary{graphs}			
\usetikzlibrary{graphdrawing}	
\usetikzlibrary{quotes}			
\usetikzlibrary{babel}          
\usetikzlibrary{hobby}
\usetikzlibrary{arrows.meta}
\usetikzlibrary{backgrounds}
\usetikzlibrary{bending}

\usegdlibrary{trees}			
\usegdlibrary{layered}			
\usegdlibrary{force}			
\usegdlibrary{circular}			
\usegdlibrary{routing}			

\usepackage{pgfplots}
\pgfplotsset{compat=1.12}

\makeatletter
\def\myfoldheight{0.5}

\def\myshapepath{
\pgfextract@process\northwest{
        \southwest\pgf@xa=\pgf@x
        \northeast
        \pgf@x=\pgf@xa
}

\pgfextract@process\southeast{
        \southwest\pgf@ya=\pgf@y
        \northeast
        \pgf@y=\pgf@ya
}

\pgfextract@process\northfold{
        \pgfpointdiff{\southwest}{\northeast}
        \northeast
        \advance\pgf@x-\myfoldheight\pgf@y
}

\pgfextract@process\eastfold{
        \pgfpointdiff{\southwest}{\northeast}
         \northeast
        \advance\pgf@y-\myfoldheight\pgf@y
}

\pgfextract@process\fold{
        \northfold\pgf@xa=\pgf@x
        \eastfold
        \pgf@x=\pgf@xa
}

        \pgfpathmoveto{\southwest}
        \pgfpathlineto{\northwest}
        \pgfpathlineto{\northfold}
        \pgfpathlineto{\eastfold}
        \pgfpathlineto{\southeast}
        \pgfpathclose
}

\def\myshapeanchorborder#1#2{
    \pgftransformreset 
    \pgf@relevantforpicturesizefalse 
    \pgfintersectionofpaths{
        \myshapepath
    }{
        \pgfpathmoveto{
            \pgfpointadd{
                \pgfpointdiff{\southwest}{\northeast}\pgf@xc=\pgf@x \advance\pgf@xc by \pgf@y 
                \pgfpointscale{
                    \pgf@xc
                }{
                    \pgfpointnormalised{
                        #2
                    }
                }
            } {
                #1
            }
        }
        \pgfpathlineto{#1}
    }
    \pgfpointintersectionsolution{1}
}

\newdimen\myshapedimenx
\newdimen\myshapedimeny

\pgfdeclareshape{file}{
    \inheritsavedanchors[from=rectangle]
    \inheritanchor[from=rectangle]{center}
    \inheritanchor[from=rectangle]{mid}
    \inheritanchor[from=rectangle]{base}

    \inheritanchor[from=rectangle]{west}
    \inheritanchor[from=rectangle]{east}
    \inheritanchor[from=rectangle]{north}
    \inheritanchor[from=rectangle]{south}

    \inheritanchor[from=rectangle]{south west}
    \inheritanchor[from=rectangle]{south east}
    \inheritanchor[from=rectangle]{north west}

    \inheritanchor[from=rectangle]{mid west}
    \inheritanchor[from=rectangle]{mid east}
    \inheritanchor[from=rectangle]{base west}
    \inheritanchor[from=rectangle]{base east}

    \backgroundpath{
       \myshapepath
     }

    \foregroundpath{
     \pgfpathmoveto{\northfold}
     \pgfpathlineto{\fold}
     \pgfpathlineto{\eastfold}
    }

  \anchorborder{%
    \pgf@xb=\pgf@x
    \pgf@yb=\pgf@y%
    \southwest%
    \pgf@xa=\pgf@x
    \pgf@ya=\pgf@y%
    \northeast%
    \advance\pgf@x by-\pgf@xa%
    \advance\pgf@y by-\pgf@ya%
    \pgf@xc=.5\pgf@x
    \pgf@yc=.5\pgf@y%
    \advance\pgf@xa by\pgf@xc
    \advance\pgf@ya by\pgf@yc%
    \edef\pgf@marshal{%
      \noexpand\pgfpointborderrectangle
      {\noexpand\pgfqpoint{\the\pgf@xb}{\the\pgf@yb}}
      {\noexpand\pgfqpoint{\the\pgf@xc}{\the\pgf@yc}}%
    }%
    \pgf@process{\pgf@marshal}%
    \advance\pgf@x by\pgf@xa%
    \advance\pgf@y by\pgf@ya%
  }

}

\makeatother

\makeatletter
\newcommand\miniscule{\@setfontsize\miniscule{4}{5}}
\makeatother

\fi

\let\svthefootnote\thefootnote
\newcommand\blankfootnote[1]{%
  \let\thefootnote\relax\footnotetext{#1}%
  \let\thefootnote\svthefootnote%
}

\usepackage{pgfmath}

\makeatletter
\newsavebox{\my@scalepar@TempBox}

\makeatother

\usepackage{color, colortbl}

\begin{document}

\title{User-Directed Loop-Transformations in Clang%
}

\author{\IEEEauthorblockN{Michael Kruse}
	\IEEEauthorblockA{\textit{Argonne Leadership Computing Facility} \\
		\textit{Argonne National Laboratory}\\
		Argonne, USA\\
		mkruse@anl.gov}
	\and
	\IEEEauthorblockN{Hal Finkel}
	\IEEEauthorblockA{\textit{Argonne Leadership Computing Facility} \\
		\textit{Argonne National Laboratory}\\
		Argonne, USA\\
		hfinkel@anl.gov}
	\and
}

\maketitle

\begin{abstract}
Directives for the compiler such as pragmas can help programmers to separate an algorithm's semantics from its optimization. This keeps the code understandable and easier to optimize for different platforms. Simple transformations such as loop unrolling are already implemented in most mainstream compilers.
We recently submitted a proposal to add generalized loop transformations to the OpenMP standard. We are also working on an implementation in LLVM/Clang/Polly to show its feasibility and usefulness. The current prototype allows applying patterns common to matrix-matrix multiplication optimizations.
\end{abstract}



\definecolor{darkbrown}{rgb}{0.4, 0.26, 0.13}
\definecolor{dartmouthgreen}{rgb}{0.05, 0.5, 0.06}
\definecolor{crimsonglory}{rgb}{0.75, 0.0, 0.2}
\definecolor{chromeyellow}{rgb}{1.0, 0.65, 0.0}
\definecolor{amethyst}{rgb}{0.6, 0.4, 0.8}
\definecolor{darkgoldenrod}{rgb}{0.72, 0.53, 0.04}
\definecolor{darkmagenta}{rgb}{0.55, 0.0, 0.55}
\definecolor{babyblueeyes}{rgb}{0.63, 0.79, 0.95}
\definecolor{blizzardblue}{rgb}{0.67, 0.9, 0.93}
\definecolor{lightcyan}{rgb}{0.88, 1.0, 1.0}

\newcommand\pragma[1]{\texttt{\#pragma~#1}}
\newcommand\pragmaomp[1]{\texttt{\#pragma~omp~#1}}
\newcommand*\joop[1]{\texttt{#1}}
\newcommand*\data[1]{\texttt{#1}}
\newcommand*\sect[1]{\texttt{#1}}
\newcommand\syntax[1]{\textbf{\texttt{#1}}}
\newcommand\placeholder[1]{\textrm{\textit{#1}}}

\section{Motivation}\label{sct:motivation}


Almost all processor time is spent in some kind of loop, and as a result, loops are a primary target for program optimization efforts.
Changing the code directly comes with the disadvantage of making the code much less maintainable. 
That is, it makes reading and understanding the code more difficult, bugs occur easier and making semantic changes that would be simple in an unoptimized version might require large changes in the optimized variant. 
Porting to a new system architecture may require redoing the entire optimization effort and maintain several copies of the same code each optimized for a different target, even if the optimization does not get down to the assembly level. 
Understandably, this is usually only done for the most performance-critical functions of a code base, if at all.

For this reason, mainstream compilers implement pragma directives with the goal of separating the code's semantics from its optimization. 
That is, the code should compute the same result if the directives are not present.
For instance, pragmas defined by the OpenMP standard~\cite{openmp} and supported by most of the mainstream compilers will parallelize code to run using multiple threads on the same machine. 
An alternative is to use platform-specific thread libraries such as pthreads. 
OpenMP also defines directives for vectorization and accelerator offloading. 
Besides OpenMP, there are a few more sets of pragma directives, such as OpenACC~\cite{openacc}, OmpSs~\cite{openss}, OpenHMPP~\cite{openhmpp}, OpenStream~\cite{openstream}, etc., with limited compiler support.


Besides the standardized pragmas, most compilers implement pragmas specific to their implementation.
\Cref{tbl:proprietarypragmas} shows a selection of pragmas supported by popular compilers.
By their nature, support and syntax varies heavily between compilers.
Only \pragma{unroll} has broad support.
\pragma{ivdep} made popular by icc and Cray to help vectorization is mimicked by other compilers as well, but with different interpretations of its meaning.
However, no compiler allows applying multiple transformations on a single loop systematically.

\begin{table*}
\begin{tabular}{rll}
	Transformation                                            & Syntax                                                       & Compiler Support                                                                            \\ 
\cmidrule(r){1-1} \cmidrule(lr){2-2} \cmidrule(l){3-3}
\rowcolor{lightcyan}	Threading                                                 & \pragmaomp{parallel for}                                     & OpenMP~\cite{openmp}                                                                \\
\rowcolor{lightcyan}	                                                          & \pragma{loop(hint\_parallel(0))}                             & msvc~\cite{msvcloop}                                                                \\
\rowcolor{lightcyan}	                                                          & \pragma{parallel}                                            & icc~\cite{iccmanual}                                                                \\
\rowcolor{lightcyan}	                                                          & \pragma{concur}                                              & PGI~\cite{pgi-pragmas} \\
	Offloading                                                & \pragmaomp{target}                                           & OpenMP~\cite{openmp}                                                                \\
	                                                          & \pragma{acc kernels}                                         & OpenACC~\cite{openacc}                                                              \\
	                                                          & \pragma{offload}                                             & icc~\cite{iccmanual}                                                                \\
\rowcolor{lightcyan}	Unrolling                                                 & \pragma{unroll}                                              & \minibox{icc~\cite{iccmanual}, xlc~\cite{xlcmanual}, clang~\cite{clangattributes}, Oracle~\cite{suncc-pragmas},\\ PGI~\cite{pgi-pragmas}, SGI~\cite{sgi-lno}, HP~\cite{acc}}  \\
\rowcolor{lightcyan}	                                                          & \pragma{clang loop unroll(enable)}                           & clang~\cite{clangextensions}                                                        \\
\rowcolor{lightcyan}	                                                          & \pragma{GCC unroll \placeholder{n}}                          & gcc~\cite{gccpragmas}                                                               \\
\rowcolor{lightcyan}	                                                          & \pragma{\_CRI unroll}                                        & Cray~\cite{craycc-pragmas} \\
	Unroll-and-jam                                            & \pragma{unroll\_and\_jam}                                    & icc~\cite{iccmanual}, clang~\cite{clang-unrollandjam}                                                                \\
					                                          & \pragma{unroll}                                              & SGI~\cite{sgi-lno} \\
	                                                          & \pragma{unrollandfuse}                                       & xlc~\cite{xlcmanual}                                                                \\
	                                                          & \pragma{stream\_unroll}                                      & xlc~\cite{xlcmanual}                                                                \\
\rowcolor{lightcyan}	Loop fusion                                               & \pragma{nofusion}                                            & icc~\cite{iccmanual}                                                                \\
\rowcolor{lightcyan}									                          & \pragma{fuse}                                                & SGI~\cite{sgi-lno}\\
\rowcolor{lightcyan}									                          & \pragma{\_CRI fusion}                                        & Cray~\cite{craycc-pragmas} \\
	Loop distribution                                         & \pragma{distribute\_point}                                   & icc~\cite{iccmanual}                                                                \\
	                                                          & \pragma{clang loop distribute(enable)}                       & clang\cite{clangextensions}                                                         \\
	                                                          & \pragma{fission}                                             & SGI~\cite{sgi-lno}\\
	                                                          & \pragma{\_CRI nofission}                                     & Cray~\cite{craycc-pragmas}\\	                                                          									
\rowcolor{lightcyan}	Loop blocking                                             & \pragma{block\_loop}                                         & xlc~\cite{xlcmanual}                                                                \\
\rowcolor{lightcyan}                                                              & \pragma{blocking size}                                       & SGI~\cite{sgi-lno} \\
\rowcolor{lightcyan}                                                              & \pragma{\_CRI blockingsize}                                  & Cray~\cite{craycc-pragmas} \\
	Vectorization                                             & \pragmaomp{simd}                                             & OpenMP~\cite{openmp}                                                                \\
	                                                          & \pragma{simd}                                                & icc~\cite{iccmanual}                                                                \\
	                                                          & \pragma{vector}                                              & icc~\cite{iccmanual}, PGI~\cite{pgi-pragmas}                                                                \\
	                                                          & \pragma{loop(no\_vector)}                                    & msvc~\cite{msvcloop}                                                                \\
	                                                          & \pragma{clang loop vectorize(enable)}                        & clang~\cite{clangextensions,clangvectorizers}                                       \\
\rowcolor{lightcyan}	Interleaving                                              & \pragma{clang loop interleave(enable)}                       & clang~\cite{clangextensions,clangvectorizers}                                       \\
	Software pipelining                                       & \pragma{swp}                                                 & icc~\cite{iccmanual}                                                                \\
                                                              & \pragma{pipeloop}                                            & Oracle~\cite{suncc-pragmas} \\
\rowcolor{lightcyan}Loop name                                                     & \pragma{loopid}                                              & xlc~\cite{xlcmanual}\\
Loop versioning                                               & \pragma{altcode}                                             & PGI~\cite{pgi-pragmas} \\	
\rowcolor{lightcyan}Loop-invariant code motion                                    & \pragma{noinvarif}                                           & PGI~\cite{pgi-pragmas} \\	
Prefetching                                                   & \pragma{mem prefetch}                                        & PGI~\cite{pgi-pragmas}\\
                                                              & \pragma{prefetch}                                            & SGI~\cite{sgi-lno}\\
\rowcolor{lightcyan}Interchange                                                   & \pragma{interchange}                                         & SGI~\cite{sgi-lno}\\
\rowcolor{lightcyan}                                                              & \pragma{\_CRI interchange}                                   & Cray~\cite{craycc-pragmas}\\
If-conversion                                                 & \pragma{IF\_CONVERT}                                         & HP~\cite{acc} \\
\rowcolor{lightcyan}Collapse loops                                                & \pragma{\_CRI collapse}                                      & Cray~\cite{craycc-pragmas}\\
Assume iteration independence                                 & \pragma{ivdep}                                               & icc~\cite{iccmanual}, PGI~\cite{pgi-pragmas}, SGI~\cite{sgi-lno},                   \\
	                                                          & \pragma{GCC ivdep}                                           & gcc~\cite{gccpragmas}                                                               \\
	                                                          & \pragma{loop(ivdep)}                                         & msvc~\cite{msvcloop}                                                                \\
	                                                          & \pragma{nomemorydepend}                                      & Oracle~\cite{suncc-pragmas}\\
	                                                          & \pragma{nodepchk}                                            & PGI~\cite{pgi-pragmas}, HP~\cite{acc} \\
\rowcolor{lightcyan}Iteration count estimation                                    & \pragma{loop\_count(\placeholder{n})}                        & icc~\cite{iccmanual}     \\
\end{tabular}\centering
\caption{Loop pragmas and the compilers which support them}\label{tbl:proprietarypragmas}
\end{table*}

In addition to straightforward trial-and-error execution time optimization, code transformation pragmas can be useful for machine-learning assisted autotuning.
The OpenMP approach is to make the programmer responsible for the semantic correctness of the transformation.
This unfortunately makes it hard for an autotuner which only measures the timing difference without understanding the code. 
Such an autotuner would therefore likely suggest transformations that make the program return wrong results or crash.
Assistance by the compiler which understands the semantics and thus can either refuse to apply a transformation or insert fallback code (\emph{code versioning}) that is executed instead if the transformation is unsafe enables loop autotuning.
Even for programmer-controlled programs, warnings by the compiler about transformations that might change the program's result can be helpful.

In summary, pragma directives for code transformations are useful for assisting program optimization and are already widely used.
However, outside of OpenMP, these cannot be used for portable code since compilers disagree on syntax, semantics, and only support a subset of transformations.

Our contributions for improving the usability of user-directed loop transformations are
\begin{itemize}
\item the idea of making user-directed loop transformations arbitrarily composable,
\item an effort to standardize loop-transformation pragmas in OpenMP~\cite{iwomp18-pragmas}, and
\item a prototype implementation using Clang and Polly that implements additional loop-transformation pragmas.
\end{itemize}

\section{Pragma Directives in Clang}

Clang in the current version (7.0) already supports the following pragma directives:
\begin{itemize}
\item Thread-parallelism: \pragmaomp{parallel}, \pragmaomp{task}, etc.
\item Accelerator offloading: \pragmaomp{target}
\item \pragma{clang loop unroll} (or \pragma{unroll})
\item \pragma{unroll\_and\_jam}
\item \pragma{clang loop distribute(enable)}
\item \pragma{clang loop vectorize(enable)} (or \pragmaomp{simd})
\item \pragma{clang loop interleave(enable)}
\end{itemize}

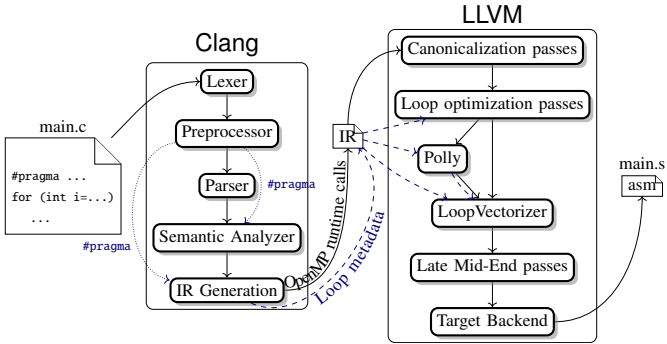
\begin{figure}
	\resizebox{\linewidth}{!}{%
		\begin{tikzpicture}%
		\tikzset{tight/.style={inner sep=0pt,outer sep=0pt,minimum size=0pt}}
		\tikzset{node/.style={draw,fill=white,line width=1.2pt,rounded corners,drop shadow}}
		\tikzset{supernode/.style={subgraph text none,draw,rounded corners}}
		\tikzset{edge/.style={->}}
		\graph[layered layout,edges={edge,rounded corners},level sep=5mm,sibling sep=10mm]{
			c[as={\minibox{\\\texttt{\#pragma \dots}\\\texttt{for (int i=...)}\\\hspace*{4mm}\dots}},grow=right,draw,shape=file,fill=white,label={main.c}];
			ir[as={IR},shape=file,draw,fill=white,nudge=(up:10mm)];
			asm[as={asm},shape=file,draw,fill=white,label={main.s}];
			
			clang [subgraph text none,label={[font=\Large\sffamily]above:Clang}] // [sibling sep=2mm,grow=down,layered layout] {
				lexer [as={Lexer},node];
				parser [as={Parser},node,grow=down];
				preprocessor [as={Preprocessor},node];
				sema [as={Semantic Analyzer},node];
				codegen [as={IR Generation},node];
				lexer->preprocessor->parser->sema->codegen;
			};
			llvm [subgraph text none,label={[font=\Large\sffamily]above:LLVM}] // [sibling sep=2mm,grow=down,layered layout] {
				canonicalization [as={Canonicalization passes},node];
				loopopts [as={Loop optimization passes},node];
				polly [as={Polly},node];
				vectorization [as={LoopVectorizer},node];
				latepasses [as={Late Mid-End passes},node];
				backend [as={Target Backend},node];
				canonicalization->loopopts->vectorization->latepasses->backend;
				loopopts->polly->vectorization;
			};
			c->[in=180]lexer;
			codegen->[out=0,in=-90,postaction={decorate,decoration={text along path,text={OpenMP runtime calls},raise=0.3ex}}]ir;
			ir->[out=90,in=180]canonicalization;
			backend->[out=0,in=-90]asm;
		};
		\begin{pgfonlayer}{background}
		\node[tight,fit={(clang)},supernode]{};
		\node[tight,fit={(llvm)},supernode]{};
		\end{pgfonlayer}
		\path (preprocessor) edge[edge,densely dotted,bend left=50,draw=blue!50!black] node[midway,right,font=\ttfamily,blue!50!black] {\#pragma} (sema);
		\path (preprocessor) edge[edge,densely dotted,bend right=80,draw=blue!50!black] node[pos=0.7,left,font=\ttfamily,blue!50!black] {\#pragma} (codegen);
		\path (codegen) edge[edge,dashed,bend right=80,draw=blue!50!black,postaction={decorate,decoration={text along path,text={|\color{blue!50!black}|Loop metadata},raise=-1.7ex,pre=moveto,pre length=13mm}}] (ir);
		\path (ir) edge[edge,dashed,draw=blue!50!black] (loopopts);
		\path (ir) edge[edge,dashed,draw=blue!50!black] (polly);
		\path (ir) edge[edge,dashed,draw=blue!50!black,bend right=10] (vectorization);
		\path (polly) edge[edge,dashed,draw=blue!50!black,bend right=10] (vectorization);
		\end{tikzpicture}%
	}
	\caption{Clang/LLVM compiler pipeline}\label{fig:pipeline}
\end{figure}

\subsection{Front-End}

Clang's current architecture (shown in \cref{fig:pipeline}) has two places where code transformations occur:
\begin{enumerate}
\item OpenMP (except \pragmaomp{simd}) is implemented at the front-end level:
      The generated LLVM-IR contains calls to the OpenMP runtime.
\item Compiler-driven optimizations are implemented in the mid-end:
      A set of transformation passes that each consume LLVM-IR with loops and output transformed IR, but metadata attached to loops can influence the passes' decisions.
\end{enumerate}

This split unfortunately means that OpenMP-parallel loops are opaque to the LLVM passes further down the pipeline.
Also, loops that are the result of other transformations (e.g. loop distribution) cannot be parallelized this way because the parallelization must have happened earlier in the front-end.

Multiple groups are working on improving the situation by adding parallel semantics to the IR specification~\cite{tian17,tapir}.
These and other approaches have been presented on LLVM's mailing list~\cite{pir_llvmdev17,pir_llvm18} or its conferences~\cite{pir_llvm16,pir_eurollvm18}.

\subsection{Mid-End}

Any transformation pragma not handled in the front-end, including \pragmaomp{simd}, is lowered to loop metadata. 
Metadata is a mechanism for annotating LLVM IR instructions with additional information, such as debug info.
Passes can look up these annotations and change their behavior accordingly.
An overview of loop transformation passes and the metadata they look for is shown in \cref{tbl:passes}.

\begin{table}
\begin{tabular}{ll}
	                                                                                  LLVM Pass & Metadata                                    \\
\cmidrule(r){1-1} \cmidrule(l){2-2}
	                                                              (Simple-)LoopUnswitch         & \emph{none}                                 \\
	                                                                          LoopIdiom         & \emph{none}                                 \\
	                                                                       LoopDeletion         & \emph{none}                                 \\
	                                                                    LoopInterchange$^{*}$   & \emph{none}                                 \\                
                                                                       SimpleLoopUnroll         & \texttt{llvm.loop.unroll.*}                 \\
	                                                                         LoopReroll$^{*}$   & \emph{none}                                 \\
	                                                                 LoopVersioningLICM$^{+*}$  & \texttt{llvm.loop.licm\_versioning.disable} \\
	                                                                     LoopDistribute$^{+}$   & \texttt{llvm.loop.distribute.enable}        \\
                                                           \multirow{3}{*}{LoopVectorize$^{+}$} & \texttt{llvm.loop.vectorize.*}              \\
                                                                                                & \texttt{llvm.loop.interleave.count}         \\
                                                                                                & \texttt{llvm.loop.isvectorized}             \\
                                                                    LoopLoadElimination$^{+}$   & \emph{none}                                 \\                     
	                                                                   LoopUnrollAndJam$^{*}$   & \texttt{llvm.loop.unroll\_and\_jam.*}       \\
                                                                             LoopUnroll         & \texttt{llvm.loop.unroll.*}                 \\[1.5ex]
                                                                                 \emph{various} & \texttt{llvm.mem.parallel\_loop\_access}    \\
\end{tabular}\centering
\caption{Loop transformation passes in the pipeline order and the metadata they process. 
Loop passes that canonicalize or only modify/move instructions without changing the loop structure are not included.
Passes marked as $^{*}$ are not added to any default pipeline (such as \texttt{-O3}).  
Passes with a $^{+}$ marker can do code versioning.
}
\label{tbl:passes}
\end{table}

Some of the passes must be enabled explicitly, even when using the higher optimization level.
For instance, the switch \textinline{-enable-unroll-and-jam} adds the LoopUnrollAndJam pass to the pipeline.
When using clang, \textinline{-mllvm -enable-unroll-and-jam} has to be used.
When not in the pass pipeline, any \pragma{unroll\_and\_jam} will be silently ignored.

For each function, the transformations are executed in the order of the passes in the pipeline.
This implicitly defines order in case multiple pragmas are defined on the same loop.
For instance, 
\begin{minted}{c}
#pragma clang loop vectorize(enable) distribute(enable)
\end{minted}
 will first try to distribute a loop, then vectorize the output loops.
The order of the directives and/or pragmas is irrelevant.
With this design it is not possible to apply transformations in any other order, similar to the transformation of OpenMP in the front-end. 
For instance, it is not possible to distribute a loop and then recognize a memcpy in one of the resulting loops using LoopIdiom.
It is also not possible to apply a transformation multiple times, unless the responsible pass is scheduled multiple times as well.

Ideally, the pass structure of the mid-end should be an implementation detail, but as we have seen, it is visible through the execution order of the pragmas. 
Implementation details can change between versions, and even the optimization level (\texttt{-O0}, \texttt{-O1}, \dots). 

Loop passes can use LoopVersioning to get a clone of the instructions and control flow they want to transform.
That is, each loop transformation makes its own copy of the code, with a potential code growth that is exponential in the number of passes since the versioned fallback code may still be optimized by further loop passes. 
The pass pipeline has up to 4 passes which potentially version, leading to up to 16 variants of the same code.
In addition, many versioning conditions, such as checking for array aliasing, will be similar or equal for each versioning.


\section{Extending Loop-Transformations Pragmas}

We are working on improving support for user-directed code transformations in Clang, especially loop transformations.
This includes addressing the shortcomings of the existing infrastructure described in the previous section and entirely new transformations.

If we only improve code transformations in Clang, the problem of the non-portability like for any of the pragmas in \cref{tbl:proprietarypragmas} remains.
For this reason, we submitted a proposal for inclusion into the OpenMP standard~\cite{iwomp18-pragmas}. 
We hope that this will improve the adoption of extended pragmas.

No discussion about an OpenMP standardization has started yet, hence in this paper we only describe our prototype implementation in Clang.
In particular, the pragma syntax is different because the \pragmaomp{} keyword is reserved for standardized OpenMP directives.

\subsection{Transformation Order}

The transformation order of the current pragmas is basically undefined.
For the limited set of pragmas available, transformations are either commutative or a different order makes little sense.
However, for additional transformations, the transformation order may become important.

The idea is that a pragma applies to the code that follows.
That is, if the next line is a canonical for-loop, then that loop is transformed.
If the next line is another pragma, then the output of that pragma is transformed.
This means that the example
\begin{minted}{c}
#pragma clang loop reverse
#pragma clang loop unroll factor(2)
for (int i = 0; i < 128; i+=1)
  Stmt(i);
\end{minted}
will be partially unrolled by a factor of 2, then reversed.
This will yield code comparable to
\begin{minted}{c}
for (int i = 126; i >= 0; i-=2) {
  Stmt(i);
  Stmt(i+1);
}
\end{minted}
instead of 
\begin{minted}{c}
for (int i = 127; i >= 0; i-=2) {
  Stmt(i);
  Stmt(i-1);
}
\end{minted}
that would be the result if the transformations were applied in the reverse order.

\subsection{Followup-Attributes}

The metadata from \cref{tbl:passes} is used as a bag of attributes to a loop.
By definition, their order is unimportant and cannot be used to pass ordered transformations.

In a first RFC on the LLVM mailing list~\cite{rfc_loopmetadata}, we suggested to add a function-wide (ordered) list of transformations of any loop in the function.
While this works and our current prototype uses it, it has some problems.
First, the function inliner needs to merge such lists.
Second, it requires a reference to the loop to transform. 
Unfortunately, due to the nature of IR metadata, the \enquote{loop id} currently used by LLVM changes whenever the loop's attribute list is changed\footnote{For instance, LoopVersioningLICM adds \cinline{llvm.loop.licm_versioning.disable} to indicate that it does not have to run on a loop again.}.
Third, the same \enquote{loop id} can be assigned to multiple loops. 
Naive cloning of code (like LoopVersioning does) will reuse the same \enquote{loop id} for a copied loop. 
There are even regression tests for multiple loops with the same \enquote{id}.
Fourth, transformation passes need to search the entire list for transformations they can apply and ensure that no other transformation is applied before it.

Our eventual approach~\cite{review_loopmetadata} is more compatible with \enquote{loop ids} as bag of attributes by specifying new attributes.
Every transformation gets \emph{followup-attribute lists}. 
The list defines the bag of attributes a result loop will have.  
For instance, the \cinline{llvm.loop.unroll.followup_unrolled} attribute contains the metadata for the (partially) unrolled output loop.
If not specified, the pass can automatically define the attributes; in case of LoopUnroll, it uses \cinline{llvm.loop.unroll.disable} to inhibit further unrolling.

Transformations can also have multiple output loops. 
In case of partial unrolling, there can be an epilogue for iterations that do not fill up the unroll factor. 
Its attributes can be defined using the \cinline{llvm.loop.unroll.followup_remainder} attribute.
Typically, this loop is completely unrolled such there is no loop the attributes can be assigned to.

If a transformation cannot be applied for any reason, it is straightforward to also ignore the followup-attributes as they are meant for the transformed loop which does not exist. 
We decided against applying some followups even if a transformation failed (such as applying \cinline{followup_unrolled} even if the partial unrolling failed) or add an additional \enquote{backup} attribute list.
This would significantly increase the systems complexity and the expected reaction is to fix the input code and/or transformation instead of specifying another chain of transformations.

Instead, if the transformation was explicitly requested by the programmer (which we call \enquote{forced}), the compiler should emit a warning that something did not apply as expected.
In our proposal, this is done by a new WarnMissedTransformationsPass which is inserted into the pass manager after all transformation pass have run. 
Hence, in contrast to LLVM's current approach, it will even emit a warning if the responsible pass is not even in the pipeline.
This unfortunately also means that the WarnMissedTransformations pass needs to understand all transformation metadata.

The advantage to this approach is that is more compatible with the current implementation. 
If followup-attributes are not used, the behavior remains the same, even without mitigations such as IR conversions through AutoUpgrade.

\subsection{Syntax}

Unfortunately, the existing \pragma{clang loop} syntax already has a de-facto defined order, which is the order in \cref{tbl:passes}.
To maintain compatibility, we have to introduce a distinguishable new syntax.
When exclusively using the old syntax, clang has to emit the order that original pipeline would have.
To avoid confusion, we disallow mixing old and new syntax.

The old syntax is
\begin{flushleft}
\syntax{\#pragma clang loop} \placeholder{transformation}\syntax{(}\placeholder{option}\syntax{)} \dots
\end{flushleft}
and allows multiple \placeholder{transformation}\syntax{(}\placeholder{option}\syntax{)} on each line. 
Multiple options for the same transformation can be specified by using multiple variants of \placeholder{transformation}.
For instance, \cinline{vectorize(assume_safety) vectorize_width(4)} tells the compiler to skip semantic legality checks and use an SIMD width of 4.
More complicated transformations such as tiling can have many options, which makes this syntax impractical.
It is also ambiguous whether a transformation should be applied multiple times or whether only an option is set.

Since our goal is an inclusion into the OpenMP standard, we use its directive-clause syntax:
\begin{flushleft}
\syntax{\#pragma clang loop} \placeholder{transformation} \placeholder{switch} \placeholder{option}\syntax{(}\placeholder{argument}\syntax{)} \dots
\end{flushleft}
The old and new syntax are distinguishable by the option in parenthesis immediately following the transformation keyword.

Interestingly, Clang's mechanisms for \cinline{#pragma clang loop} and OpenMP are quite different.
The prototype implementations is oriented towards the \cinline{#pragma clang loop} route, because it is also used for the preexisting loop transformation pragmas and in contrast to the OpenMP path, is less invasive.

\subsubsection{Preprocessor}

The clang-loop pragma handler takes the tokens of each clause (\placeholder{transformation}\syntax{(}\placeholder{option}\syntax{)}), wraps each of them into a \cinline{annot_pragma_loop_hint} token and pushes that back into the token stream.
For out new syntax we introduce a new token \cinline{annot_pragma_loop_transform} which contains the entire line of tokens of the pragma instead of individual clauses. 
The OpenMP pragma preprocessor also takes the entire line of tokens, encloses them between a \cinline{annot_pragma_openmp} and \cinline{annot_pragma_openmp_end} token, and pushes them all back to into the token stream.

This step is necessary because the tokens might be inside a \cinline{_Pragma("...")} of a macro. 
The preprocessor might therefore duplicate the tokens wherever the macro is expanded.
The special tokens act as indicators for the parser that such a directive has been inserted.

\subsubsection{Parser}

The parser will handle the indicator tokens at expected positions. 
If an \cinline{annot_pragma_loop_hint} is encountered, the transformation and its option are parsed, and the result used to initialize an attribute of type LoopHintAttr.
Since the data a LoopHintAttr can store is limited and unspecific, we introduce a separate attribute for each of our new transformations when encountering a \cinline{annot_pragma_loop_transform}.

The OpenMP parser on the other hand does not create attributes, but uses the Sema object (calling its \texttt{ActOn...} methods), the semantic analyzer which also creates the AST.
For parsing expressions according to the language rules, its tokens must be on the main token stack. 
For \cinline{annot_pragma_loop_*} this means that the wrapped tokens have to be pushed back to the stack before the expression parser is invoked.
This is easier for the OpenMP pragma tokens which are already on the stack, terminated by a \cinline{annot_pragma_openmp_end} token that inhibits the expression parser from consuming tokens that do not belong to the pragma.

\subsubsection{Semantic Analysis}

The semantic analyzer is responsible for building the abstract syntax tree. 
For the LoopHintAttr and other transformation attributes, it creates an AttributedStmt as parent of the node that represents the loop that is annotated.
It also checks the semantic correctness of the attributes, for instance, it emits warnings when the same \cinline{annot_pragma_loop_hint} clause is used more than once.

In the case of OpenMP, the pragmas are more invasive.
Every OpenMP-annotated loop is nested into a CapturedStmt region which is handled like an outlined function, even for \pragmaomp{simd}.
Moreover, most OpenMP directives have their own type of AST node (in addition to CapturedStmt). 
That is, the AST will look differently compared to if OpenMP was disabled.

\subsubsection{Code Generation}

While Clang generates LLVM-IR for a loop, it also collects loop attributes.
After the IR of the loop skeleton (loop header, latch, etc.) is complete, Clang sets the loop's metadata (see \cref{tbl:passes}).
Of course, loops can be nested and hence there is a stack if loop attributes called LoopInfoStack.

Since OpenMP is handled in the front-end, there is much more to do. 
Depending on the directive, a CapturedStmt is either outlined in a separate function or into the same function.
In the former case, a call to the OpenMP runtime (libomp) with a pointer to the outlined function.
In other cases, the captured-but-expanded-inline loop body will be simplified again in the mid-end.
For the \pragmaomp{simd} directive, the loop will receive a meta annotation just as in the \pragma{clang loop vectorize} case.

\subsection{Composibility}

Our goal is that the loops resulting from a transformation can again be transformed using a pragma.
For clang, there is the problem that thread-parallelism is handled in the front-end and LLVM-IR has no semantics for parallelism.
That is, its pragmas cannot apply on loops that are only created in the mid-end, nor can a \pragmaomp{for} loop be processed in the mid-end.
As a temporarily solution, we think about adding non-OpenMP pragmas for thread-parallelism that are handled in the mid-end, such as \cinline{#pragma clang loop thread_parallel}.

Moreover, transformations can have more than one input loop (such as OpenMP's \texttt{collapse} clause), and more than one output loop (such as strip-mining). 
To be able to refer to specific loops, we allow to assign identifiers to loop. 
A loop identifier must be unique within a function. For instance,
\begin{minted}{c}
#pragma clang loop id(i)
for (int i = 0; i < n; i += 1)
\end{minted}
assigns the identifier \texttt{i} to the loop. In case a loop is a canonical for-loop, its induction variable name might be used as a loop identifier unless it is ambiguous.

The loop identifiers can be used by transformations to refer to loops that are not on the next line.
For instance, unroll-and-jam requires a loop to be unrolled and one to be jammed. In the following example, the loop \texttt{i} would be unrolled-and-jammed into the \texttt{k}-loop (instead the \texttt{j}-loop which would yield two \texttt{k}-loops inside it).
\begin{minted}{c}
#pragma clang loop(i,k) unroll_and_jam factor(2)
for (int i = 0; i < n; i += 1)
  for (int j = 0; j < n; j += 1)
    for (int k = 0; k < n; k += 1)
\end{minted}

In case there are multiple output loops, the pragma can define the identifiers of the those:
\begin{minted}{c}
#pragma clang loop stripmine size(4) pit_id(i1) strip_id(i2)
for (int i = 0; i < 128; i += 1)
\end{minted}
Its result is equivalent to as if the programmer had written the following.
\begin{minted}{c}
#pragma clang loop id(i1)
for (int i1 = 0; i1 < 128; i1 += 4)
  #pragma clang loop id(i2)
  for (int i2 = i1; i2 < i1+4; i2 += 1)
\end{minted}

Some transformations can be understood as syntactic sugar for other transformations.
For instance, the aforementioned unroll-and-jam can also be expressed as an unroll followed be (one or multiple) loop fusions.
Standard loop tiling is nothing else than strip-mining of each loop, then permute the loop nest order such that the strip loops become the interior loops.


\subsection{Additional Transformations}\label{sct:transformations}

Some new transformations have already been mentioned in the examples for illustration purposes.
Many more are possible, see \cref{tbl:proprietarypragmas} and \cite{iwomp18-pragmas} for ideas.

In our prototype, we started implementing a limited set that are of immediate interest for us, mainly optimizing a matrix-matrix multiplication as shown in \cref{sct:matmul}.
In addition to \pragma{clang loop id}, we implemented the following pragmas.


\subsubsection{Loop Reversal}

Invert the iteration order of a loop. I.e.
\begin{minted}{c}
#pragma clang loop reverse
for (int i = 0; i < n; i+=1)
\end{minted}
is transformed into
\begin{minted}{c}
for (int i = n-1; i >= 0; i-=1)
\end{minted}

This was the first transformation we implemented because it is one of the simplest: Exactly one input and output loop.

\subsubsection{Loop Interchange}

Permute the order of perfectly nested loops. For instance, the result of
\begin{minted}{c}
#pragma clang loop(i,j) interchange permutation(j,i)
for (int i = 0; i < n; i+=1)
  for (int j = 0; j < m; j+=1)
\end{minted}
is
\begin{minted}{c}
for (int j = 0; j < m; j+=1)
  for (int i = 0; i < n; i+=1)
\end{minted}

The order of more than two loops can be altered by explicitly specifying the permutation.

\subsubsection{Tiling}

Tiling is a technique to improve temporal locality of accesses, especially of stencils. The example
\begin{minted}{c}
#pragma clang loop(i,j) tile sizes(4,8)
for (int i = 0; i < n; i+=1)
  for (int j = 0; j < m; j+=1)
\end{minted}
should be transformed into the following:
\begin{minted}{c}
for (int i1 = 0; i1 < n; i1+=4)
  for (int j1 = 0; j1 < m; j1+=8)
    for (int i2 = i1; i2 < n && i2 < i1+4; i2+=1)
      for (int j2 = j2; j2 < m && j2 < j1+8; j2+=1)
\end{minted}

Any number of loops can be tiled.
Tiling just a single loop is the same as strip-mining.

\subsubsection{Array Packing}

Temporarily copies the data of a loop's working set into a new buffer.
This may improve access locality because the extracted working set fits into a cache level and/or can be prefetched.
For example, the loop nest
\begin{minted}{c}
for (int i = 0; i < n; i+=1)
  #pragma clang loop pack array(A)
  for (int j = 0; j < 32; j+=1)
    f(A[j][i], i);
\end{minted}
is transformed to something approximately equivalent of:
\begin{minted}{c}
auto Packed_A[32];
for (int i = 0; i < n; i+=1) {
  for (int _ = 0; _ < 32; _+=1) 
    Packed_A[_] = A[_][i]; // Copy-in
  for (int j = 0; j < 32; j+=1)
    f(Packed_A[j], i);
}
\end{minted}

By default, the packed array is allocated on the stack, but with the clause \cinline{allocate(malloc)}, the memory will be allocated (and free'd) on the heap.

\subsection{Polly as Loop-Transformer}

Writing a new loop transformation pass in LLVM is a significant amount of work since it works on the low-level IR.
Some components such LoopVersioning can be reused, but even the dependency analysis will probably have to be written from scratch. 
This is despite LLVM has multiple dependency analyses (AliasAnalysis, DependenceInfo, LoopAccessInfo, PolyhedralInfo), but which have all been written with specific applications in mind.

Additionally, the pass manager architecture (neither the new nor \enquote{legacy}) does not allow dynamically repeated application of transformation passes. 
This would also amplify the code blowup due to repeated code versioning.

For our prototype implementation, we are using Polly~\cite{polly} to implement the additional transformations.
Polly takes LLVM-IR code and `lifts' is into another representation --\emph{schedule trees}~\cite{verdoolaege14} -- in which loop transformations are easier to express.
To transform loops, only the schedule tree needs to be changed and Polly takes care for everything else.

Using Polly, we can implement most transformations as follows.
First, let Polly create a schedule tree for a loop nest, then iteratively apply each transformation in the metadata to the schedule tree.
For every transformation we can check whether it violates any dependencies and if violations are found, act according to a chosen policy.
When done, Polly generates LLVM-IR from the schedule tree including code versioning.

Let's consider an example on how schedule trees are transformed.
Below we see the schedule tree of a single loop with two statements: StmtA and StmtB. 
Both statements execute in the same loop.
The execution order of the loop is defined by the lexicographic ordering of the band's schedule function.
Hence, the effective execution order of all statements is: StmtA[0], StmtB[0], StmtA[1], StmtB[1], \dots.

\begin{center}
\begin{tikzpicture}%
\tikzset{edge/.style={->}}
\graph[layered layout,edges={edge,rounded corners},level sep=5mm,sibling sep=10mm]{
	root [as={\emph{Domain}: $\{$ StmtA$[i]$, StmtB$[i]$ $\}$}];
    loop [as={\emph{Band}: $\{ \text{StmtA}[i] \texttt{->} [i]; \text{StmtB}[i] \texttt{->} [i] \mid 0 \le i < n \}$}];
    seq [as={\emph{Sequence}}];
    leafA [as={StmtA$[i]$}];
    leafB [as={StmtB$[i]$}];
    root->loop->seq->{leafA, leafB};
};
\end{tikzpicture}%
\end{center}

Interchanging the band and sequence node in the schedule tree is isomorphic to a loop distribution of the source code it represents, as shown below.
Here, the sequence ensures that all instances of StmtA are executed before any instance of StmtB, but the band ordering between the statement instances is not changed.
Therefore, this tree represents the execution order StmtA[0], StmtA[1], \dots, StmtB[0], StmtB[1], \dots.

\medskip

\resizebox{\linewidth}{!}{%
\begin{tikzpicture}%
\tikzset{edge/.style={->}}
\graph[layered layout,edges={edge,rounded corners},level sep=5mm,sibling sep=10mm]{
   	droot [as={\emph{Domain}: $\{$ StmtA$[i]$, StmtB$[i]$ $\}$}];
    dseq [as={\emph{Sequence}}];
    dloopA [as={$\{ \text{StmtA}[i] \texttt{->} [i] \mid 0 \le i < n \}$}];
    dloopB [as={$\{ \text{StmtB}[i] \texttt{->} [i] \mid 0 \le i < n \}$}];
    dleafA [as={StmtA$[i]$}];
    dleafB [as={StmtB$[i]$}];
    droot->dseq->{dloopA, dloopB};
    dloopA->dleafA;
    dloopB->dleafB;
};
\end{tikzpicture}%
}

Polly's infrastructure then converts the schedule tree into an AST and then back to LLVM-IR, but with the transformation applied.
Only a single fallback copy for arbitrarily many transformations is generated, if needed at all.
Once the runtime check confirms that the preconditions making the transformation valid (e.g. no overlapping memory regions), no additional checking is required.
Another advantage of a single dedicated loop transformation pass is that the analyses, including dependency analysis, needs to happen once only, instead repeatedly for every transformation.

If desired, Polly can also apply its loop nest optimizer which utilizes a linear program solver before IR generation. 
We add artificial \emph{transformational dependencies} to ensure that user-defined transformations are not overridden, but we did not implement this is the prototype yet.

As an exception, \pragma{clang loop pack} cannot implemented using this technique as it is mainly a data layout transformation. 
Only the copy-in and copy-out to the local memory allocation modify the schedule tree; these are new statements that are inserted before, respectively after the code that uses them. 
Parts of the code already existed in Polly as part of its matrix-matrix multiplication optimization~\cite{gareev18}, but had to be generalized to arbitrary loops.
To define an index function and size of the packed array, we use the bounding box technique from~\cite{molly}.
The possibility to allocate such arrays on the heap instead on the stack (\cinline{allocate(malloc)}) has been added in a Google Summer of Code project~\cite{bonfante17}.

\section{Matrix-Matrix Multiplication}\label{sct:matmul}

We chose this matrix-matrix multiplication to illustrate the power of user-directed transformations with relatively few lines and separation of semantics and optimization through pragmas.
Matrix-matrix multiplication is one of the best studied problems for performance-optimization with many BLAS library implementations that we can assume to be close to the best attainable performance. 
Comparing to them allows estimating the gap between hand-optimized implementations and compiler-produced code.
In this paper we are not searching for the best transformation chains for arbitrary algorithms\footnote{We could only compare the pragma implementation with our manual replication of the same -- which should behave identically.}, but to show how pragmas can make implementing such algorithms easier, potentially even in those specialized libraries.

\begin{listing}
\begin{minted}{c}
#if __kabylake__
  #pragma clang loop(j2) pack array(A)
  #pragma clang loop(i1) pack array(B)
  #pragma clang loop(i1,j1,k1,i2,j2,k2) interchange \
          permutation(j1,k1,i1,j2,i2,k2)
  #pragma clang loop(i,j,k) tile sizes(96,2048,256) \
          pit_ids(i1,j1,k1) tile_ids(i2,j2,k2)
#elif __haswell__
  [...]
#endif

#pragma clang loop id(i)
for (int i = 0; i < M; i+=1)
  #pragma clang loop id(j)
  for (int j = 0; j < N; j+=1) 
    #pragma clang loop id(k)
    for (int k = 0; k < K; k+=1)
      C[i][j] += A[i][k] * B[k][j];
\end{minted}
\caption{Optimization of matrix-matrix multiplication using our proposed pragmas.
The tile sizes were derived using the analytical model in \cite{low16} for Intel's Kaby Lake architecture.}
\label{lst:gemm}
\end{listing}

Fortunately, the paper~\cite{low16} describes the common techniques for optimizing matrix-matrix multiplication such that we do not need to find the optimal transformations ourselves. 
Our version is shown in \cref{lst:gemm}, which also illustrates how different transformations can be applied for different compilation targets.
It only contains the most performance-sensitive loop. 
Most of the time, matrix-matrix multiplication is written including a statement \cinline{C[i][j] = 0} in the loop nest to clear the content the array C might have had before. 
To also optimize this formulation, we would have to loop-distribute the set-zero statement and inner reduction into two different loop nests. 
Unfortunately, our prototype does not support loop distribution yet and LLVM's LoopDistribution pass is not able to handle deeply nested loops.
With loop distribution, the set-zero loop nest could be transformed into a single \cinline{memset} call. 
Again, LLVM's LoopIdiom pass only supports innermost loops such that it would result in a loop of memsets.

\begin{listing*}
\begin{minted}[escapeinside=``]{c}
double Packed_B[256][2048];
double Packed_A[96][256];
if (`\emph{runtime check}`) {
  if (M >= 1)
    for (int c0 = 0; c0 <= floord(N - 1, 2048); c0 += 1)    // Loop j1
      for (int c1 = 0; c1 <= floord(K - 1, 256); c1 += 1) { // Loop k1
      
        // Copy-in: B -> Packed_B
        for (int c4 = 0; c4 <= min(2047, N - 2048 * c0 - 1); c4 += 1)
          for (int c5 = 0; c5 <= min(255, K - 256 * c1 - 1); c5 += 1)
            Packed_B[c4][c5] = B[256 * c1 + c5][2048 * c0 + c4];

        for (int c2 = 0; c2 <= floord(M - 1, 96); c2 += 1) { // Loop i1
        
          // Copy-in: A -> Packed_A
          for (int c6 = 0; c6 <= min(95, M - 96 * c2 - 1); c6 += 1)
            for (int c7 = 0; c7 <= min(255, K - 256 * c1 - 1); c7 += 1)
              Packed_A[c6][c7] = A[96 * c2 + c6][256 * c1 + c7]; 
               
          for (int c3 = 0; c3 <= min(2047, N - 2048 * c0 - 1); c3 += 1)   // Loop j2
            for (int c4 = 0; c4 <= min(95, M - 96 * c2 - 1); c4 += 1)     // Loop i2
              for (int c5 = 0; c5 <= min(255, K - 256 * c1 - 1); c5 += 1) // Loop k2
                C[96 * c2 + c4][2048 * c0 + c3] += Packed_A[c4][c5] * Packed_B[c3][c5];
        }
     }
} else {  /* original code */ }
\end{minted}
\caption{Transformed loop nest of \cref{lst:gemm}; this is the AST as emitted by Polly (\texttt{-mllvm -debug-only=polly-ast}) modified for readability.}
\label{lst:gemmast}
\end{listing*}

Polly's main output is LLVM's intermediate representation, but it can also dump the AST representation from which the IR is generated, shown in \Cref{lst:gemmast}.  
The AST representation is one of isl's data structures.

The runtime check verifies the assumptions that must hold to make this transformation valid. 
For instance, it ensures that that arrays \texttt{A} and \texttt{B} do not overlap.

It is important that the array \texttt{Packed\_B} is transposed.
Otherwise, the accesses to are strided (not consecutive) in the innermost loop \texttt{c5} such that the processor's prefetcher will not work as efficient and the L1 cache lines are not fully used.
In our experiments, without transpose the kernel was ten times slower.

The technique in~\cite{low16} suggests that the dimensions \texttt{j2} and \texttt{i2} should be vectorized. 
In contrast to the loop \texttt{k2}, these do not carry the reduction, hence have less data-flow dependencies.
Unfortunately, LLVM's LoopVectorize currently only supports innermost loops, hence the \texttt{k2}-loop is the only one we can directly vectorize, but its heuristics decide that it is profitable without us having to add another pragma.
Extending LoopVectorize to more than innermost loops is work in progress~\cite{vplan}.
Another project addressing the issue is the Unified Region Vectorizer~\cite{rv}, which is not part of LLVM.
Polly's matrix-matrix multiplication optimization~\cite{gareev18} has a workaround that unroll-and-jams (it calls it register tiling) the loops to be vectorized and then relies on the SLPVectorizer to combine the unrolled instructions to vector instructions. 
The same unroll-and-jam is also applied on the packed arrays.
We could also try to use Polly's vector code generator (\textinline{-mllvm -polly-vectorizer}) which unfortunately also prioritizes innermost loops.

\subsection{Effectiveness}\label{evaluation}

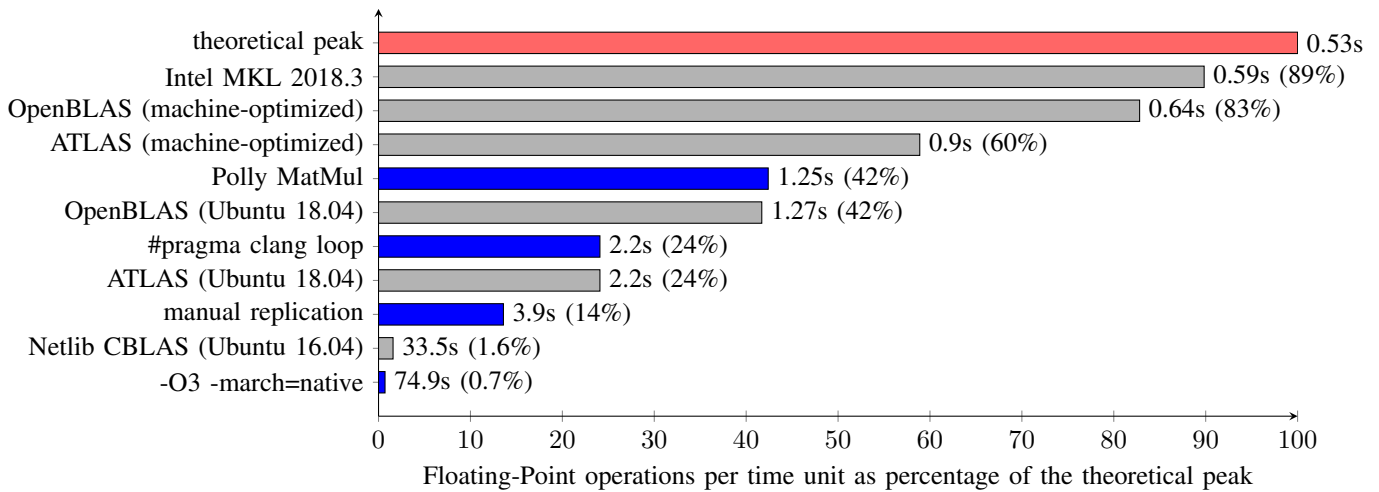
\begin{figure*}
\begin{tikzpicture}
\begin{axis}[
xbar,width=138mm,height=70mm,bar width=1.8ex,bar shift=0pt,
ytick=data,symbolic y coords={base,cblas,manual,atlas,pragma,openblas,polly,atlas_make,openblas_spack,mkl,peak},yticklabels={{-O3 -march=native},{Netlib CBLAS (Ubuntu 16.04)},{manual replication},{ATLAS (Ubuntu 18.04)},{\#pragma clang loop},{OpenBLAS (Ubuntu 18.04)},{Polly MatMul},{ATLAS (machine-optimized)},{OpenBLAS (machine-optimized)},{Intel MKL 2018.3},{theoretical peak}},axis y line=left,enlarge y limits=0.1,           
xmin=0,xmax=100,xlabel={Floating-Point operations per time unit as percentage of the theoretical peak},axis x line=bottom,    
nodes near coords={\textcolor{black}{\pgfplotspointmeta}},point meta=explicit symbolic,nodes near coords align={horizontal},nodes near coords style={right} %
]

\addplot[draw=none,fill=none] coordinates {
(0,base) 
(0,cblas)
(0,manual)
(0,atlas) 
(0,pragma)  
(0,openblas)
(0,polly)  
(0,atlas_make)
(0,openblas_spack)
(0,mkl)
(0,peak)   
};

\addplot[draw=black,fill=gray!60] coordinates {
(1.58,cblas)  [33.5s (1.6\%)]
(24.09,atlas)   [2.2s (24\%)] 
(58.8889,atlas_make) [0.9s (60\%)] 
(41.7,openblas)[1.27s (42\%)] 
(82.81,openblas_spack)[0.64s (83\%)] 
(89.83,mkl)     [0.59s (89\%)]
};

\addplot[draw=black,fill=blue] coordinates {
(0.7076,base)   [74.9s (0.7\%)]
(13.5897,manual)  [3.9s (14\%)]
(24.09,pragma)   [2.2s (24\%)] 
(42.4,polly)    [1.25s (42\%)]
};

\addplot[draw=black,fill=red!60] coordinates {
(100,peak)    [0.53s] 
};

\end{axis}
\end{tikzpicture}
\caption{Comparison of double-precision matrix-multiplication performance on an Intel Core i7 7700HQ (Kaby Lake architecture), 2.8 Ghz, Turbo Boost off}
\label{fig:dgemm}
\end{figure*}

We tested the execution speed of \cref{lst:gemm,lst:gemmast}, and compared it compared it with other implementations such as various BLAS libraries. 
All execution times were taken for a single-thread double-precision matrix-multiplication kernel using the parameters $M=2000$, $N=2300$, $K=2600$.
The results are shown in \cref{fig:dgemm}.

The na{\"i}ve version (\cref{lst:gemm} without pragmas) compiled with Clang 7.0 executes in 75 seconds (gcc's results are similar).
With the pragma transformations manually applied (like \cref{lst:gemmast} but without \cinline{floord} calls) the execution time shrinks to 3.9 seconds.
If Polly applies these transformations as directed by the pragmas in \cref{lst:gemmast}, the runtime shrinks even more to 2.2 seconds. 
This is possible because Polly applies additional metadata that indicate that, for instance, the arrays do not alias. 
Polly's matrix-matrix multiplication recognition~\cite{gareev18} optimizes the kernel such that it runs in 1.14s, which is 42\% of the processor's theoretical floating-point limit. 
This speed should eventually also be reachable using pragmas after we improved the vectorizer situation.

By comparison, Netlib's reference BLAS implemented requires 33.5 seconds to do the multiplication.
ATLAS from the Ubuntu 18.04 (Xenial Xerus) software repository needs 2.2s for the same work, but when optimized for the target machine, only needs 0.9 seconds.
OpenBLAS, also from the Ubuntu software repository, needs 1.3 seconds, and only 0.6 seconds when compiled for the target machine.
Intel's MKL library runs in 0.59 seconds, which is impressive 89\% of the theoretical flop-limited peak performance.
ATLAS, OpenBLAS and MKL use hand-written vector kernels for the innermost tiles.

\section{Related Work}

As we have seen in \cref{tbl:proprietarypragmas}, other compilers also implement pragmas.
Most of them only allow controlling their existing loop optimization passes in the pipeline in the same manner as LLVM does.
For instance, gcc version 8.1~\cite{gccpragmas} adds support for \pragma{unroll}, but does not support any other transformation pragma although gcc has more loop transformations.

We used Polly to implement most loop transformations, but by default it tries to automatically determine which transformations are profitable using linear programming.
Apart from the equation system solver, Polly also applies tiling and a matrix-matrix multiplication optimization~\cite{gareev18} whenever possible but can only be controlled via command-line flags, not in-source annotations. 

IBM's xlf compiler has a dedicated loop transformation component, called ASTI~\cite{vivek97}.
It's data structure is the Loop Structure Graph (LSG) which shares some similarities to isl's schedule trees and might be able to carry-out xlf's supported pragmas.

Silicon Graphics also developed a compiler with a dedicated loop transformation phase called Loop Nest Optimization (LNO)~\cite{sgi-lno} which was used to implements its numerous (compared to other compilers) loop transformations.
Today, the compiler lives on in its derivatives.
One of them is the open source Open64 compiler~\cite{open64} which also contains the LNO component.

Multiple research groups already explored the composition of loop transformations, many of them based on the polyhedral model.  
The Unifying Reordering Framework~\cite{utf} describes loop transformations mathematically, including semantic legality and code generations.
The Clint~\cite{clint} tool is able to visualize multiple loop transformations.

Many source-to-source compilers can apply the loop transformations themselves and generate a new source file with the transformation baked-in. 
The instructions of which transformations to apply can be in the source file itself like in a comment of the input language (Clay~\cite{clay}, Goofi~\cite{goofi}, Orio~\cite{orio}) or like our proposal as a pragma (X-Language~\cite{xlang}, HMPP~\cite{hmpp}).
Goofi also comes with a graphical tool with a preview of the loop transformations.
The other possibility is to have the transformations in a separate file, as done by URUK~\cite{uruk} and CHiLL~\cite{chill}.
POET~\cite{poet} uses an XML-like description file that only contains the loop body code in the target language.

Halide~\cite{halide} and Tensor Comprehensions~\cite{tensorcomprehensions} are both libraries that include a compiler.
In Halide, a syntax tree is created from C++ expression templates.
In Tensor Comprehensions, the source is passed as a string which is parsed by the library.
Both libraries have objects representing the code and calling its methods transform the represented code.

Similar to the parallel extensions in the C++17~\cite{cpp17} standard library, Intel's Threading Building Blocks~\cite{tbb}, RAJA~\cite{raja} and Kokkos~\cite{kokkos} are template libraries.
The payload code is written using lambdas and an \emph{execution policy} specifies how it should be called.

Our intended use case -- autotuning loop transformations -- has also been explored by POET~\cite{poet} and Orio~\cite{orio}.
The atJIT~\cite{atjit} project is even able to use our extended transformation in Polly.  It tries out different transformations of a function within as single processes using a Just-In-Time compiler.

\section{Conclusion}

Our goal is to make more loop transformations via pragma directives available to the programmer.
Such pragmas would make applying common loop optimization technique much easier and allow better separation of a code's semantics and its optimization.

We are working on two fronts to make it happen:
First, we try to add such pragmas to the OpenMP standard~\cite{iwomp18-pragmas}.  
This would encourage any compiler that claims OpenMP-compatibility to implement them.

The second approach is to implement such pragmas in LLVM.
In this paper we presented the details of our prototype implementation using Clang to parse the pragmas and Polly to carry-out the transformations.
Experiments on matrix-matrix multiplication code show that kernels optimized using our pragmas can -- performance-wise -- be in the realm of hand-optimized BLAS libraries.

\section{Acknowledgments}

This research was supported by the Exascale Computing Project (17-SC-20-SC), a collaborative effort of two U.S. Department of Energy organizations (Office of Science and the National Nuclear Security Administration) responsible for the planning and preparation of a capable exascale ecosystem, including software, applications, hardware, advanced system engineering, and early testbed platforms, in support of the nation’s exascale computing imperative.

This research used resources of the Argonne Leadership Computing Facility, which is a DOE Office of Science User Facility supported under Contract DE-AC02-06CH11357.


\bibliographystyle{IEEEtran}
\bibliography{IEEEabrv,bibliography}

\end{document}